\documentclass[aps,prb,superscriptaddress,sd,amsmath,reprint,amssymb]{revtex4-1}
\usepackage[T1]{fontenc}

\usepackage[utf8]{inputenc}
\usepackage{graphicx}% Include figure files
\usepackage{dcolumn}% Align table columns on decimal point
\usepackage{bm}% bold math
\usepackage{color}
\usepackage{eurosym}
\usepackage{hyperref}

\begin{document}

\title{RBS/Channeling characterization of Ru(0001) and thin epitaxial Ru/Al$_2$O$_3$(0001) films.}

\author{J.E. Prieto}
\affiliation{Instituto de Qu\'{\i}mica F\'{\i}sica ``Rocasolano'', CSIC, Madrid E-28006, Spain}
\author{E.M. Trapero}
\affiliation{Instituto de Qu\'{\i}mica F\'{\i}sica ``Rocasolano'', CSIC, Madrid E-28006, Spain}
\author{P. Prieto}
\affiliation{Dpto. F\'{\i}sica Aplicada, Universidad Autónoma de Madrid, 28049 Madrid, Spain}
\author{E. García-Martín}
\affiliation{Instituto de Qu\'{\i}mica F\'{\i}sica ``Rocasolano'', CSIC, Madrid E-28006, Spain}
\author{G.D. Soria}
\affiliation{Instituto de Qu\'{\i}mica F\'{\i}sica ``Rocasolano'', CSIC, Madrid E-28006, Spain}
\author{P. Gal\'an}
\affiliation{Centro de Microan\'alisis de Materiales (CMAM), Universidad Autónoma de Madrid, 28049 Madrid}
\author{J. de la Figuera}
\affiliation{Instituto de Qu\'{\i}mica F\'{\i}sica ``Rocasolano'', CSIC, Madrid E-28006, Spain}

\date{\today}

\begin{abstract}
Thin epitaxial films of metals on insulating substrates are essential for many applications, as conducting layers, in magnetic devices or as templates for further growth. In this work, we report on the growth of epitaxial Ru films on single-crystalline Al$_2$O$_3$(0001) substrates by magnetron sputtering and their subsequent systematic characterization using Rutherford backscattering spectrometry of He ions both in random and in channeling conditions. We include results of a Ru(0001) single crystal for comparison. Analysis of channeling shows that films thicker than 35 nm grow with (0001) orientation, a well-defined epitaxial relation with the substrate and a high degree of crystal quality, comparable to the Ru(0001) single crystal. Thinner films of down to 7 nm in thickness, for which relaxation of epitaxial strain is not complete, produce a similar degree of dechanneling.
%are of similar epitaxial quality as quantified by the dechanneling they produce.
The surface of the films can be prepared in a clean and ordered state in order to allow further epitaxial growth on top.   

\end{abstract}

\maketitle

\section{Introduction}

Ruthenium substrates with the (0001) orientation are widely used in research for the growth of a variety of materials, from metals to oxides\cite{FlegePSS2018} or graphene\cite{SutterNatMat2008}. Ruthenium has some interesting characteristics: it can withstand high temperatures, it does not alloy severely with many transition metals and it is relatively difficulty of oxidize (compared to other transition metals). Metals such as Mg\cite{TirmaSS2010}, Cu\cite{GuntherPRL1995}, Ag\cite{LingPRL2004}, Au\cite{HwangPRL1991}, Pd\cite{NicolasPRL2007,SantosNJP2010}, Co\cite{FaridNJP2007}, Fe, Ni and Rh\cite{KolaczkiewiczSS1999} have been grown on Ru(0001). Among the oxides grown on Ru, there are those of 3d transition metals forming highly perfect micrometric crystals, either with the rocksalt structure such as CoO\cite{LauraPhysProc2016}, Ni$_x$Co$_{1-x}$O \cite{AnnaSciRep2019} and Ni$_x$Fe$_{1-x}$O\cite{AnnaNano2020} or with the spinel structure such as magnetite\cite{SantosJPC2009,MatteoPRB2012,SandraNano2018,SandraJCP2020}, cobalt ferrite\cite{LauraAdvMat2015} and nickel ferrite\cite{AnnaSciRep2018}, as well as rare-earth oxides such as ceria\cite{FlegeNanoscale2016} and praseodimia\cite{HockerPCCP2017}. 
Also epitaxial oxides of interest for their catalytic activity, like RuO$_2$/Ru(0001)~\cite{Nien2016}, 
or IrO$_2$(110) films on RuO$_2$(110)/Ru(0001)~\cite{Abb2018,Weber2020} have been studied and graphene on ruthenium, which can be grown using carbon dissolved in the bulk or deposited from ethylene or other hydrocarbons, has been characterized in detail by several groups\cite{McCartyCarbon2009,CuiPCCP2010,BorcaNJP2010}.

The popularity of Ru single crystal substrates for film growth has suggested the use not of bulk single crystals but of high quality crystalline Ru thin films as substrates for further growth. Successful examples of this are the use of epitaxial films of Ru on Al$_2$O$_3$(0001) for the growth of ceria\cite{SauerbreyCGD2016} or graphene\cite{SutterAPL2010}. As the etching of Ru layers is well developed due to its industrial interest\cite{RuPat}, the Ru films can be used as sacrificial layers to be removed, such as the case reported for graphene\cite{SauerbreyCGD2016}. Also graphene on thin Ru films on sapphire has been proposed for more exotic uses, such as mirrors in a neutral helium atom microscope\cite{AnemonePhD2017}.

Furthermore, ruthenium thin metal films are essential for many applications. They are used industrially as conducting layers for interconnection in integrated circuits or in magnetic devices. It is a promising candidate for interconnections replacing Cu due to the good conductivity of very thin Ru films\cite{MilosevicJAP2018} coupled to their high resistance to electromigration and stability at elevated temperatures. In magnetic devices, ruthenium films are used mainly as a way of fixing the magnetization in recording media and spin valves through their promotion of antiferromagnetic coupling between magnetic layers\cite{FullertonAPL2000}. 

The structure and morphology of Ru thin films on Al$_2$O$_3$(0001) has been studied by several authors: Yamada et al. analyzed by x-ray diffraction (XRD) films grown by pulsed laser deposition at different temperatures~\cite{YamadaJJAP2002}, Milosevich et al. analyzed sputtered films by XRD and electron micro\-scopy~\cite{MilosevicJAP2018}; this latter technique was also applied by Sutter et al~\cite{SutterAPL2010} to study morphology and structure, but a complete characterization of Ru/Al$_2$O$_3$(0001) films as a function of thickness is still missing.
In this work, we have fabricated thin Ru films of different thicknesses between 7~nm and 100~nm on Al$_2$O$_3$(0001) by magnetron sputtering and have characterized their structure by means of Rutherford backscattering spectrometry (RBS), performed both in random and in channeling geometries. This allows us to obtain information both on the crystalline structure of the films and on their thickness and composition, in addition to the structural information that can be obtained by e.g. standard XRD. For comparison, results obtained on a Ru(0001) single crystal have been included.

\section{Experimental}

Ru films have been deposited on Al$_2$O$_3$(0001) single crystal substrates by means of direct-current magnetron sputtering inside a home-made sputtering chamber with a base pressure of 8$\times$10$^{-6}$ mbar.
Magnetron powers used were 20 W for the Ru films with the smallest thicknesses (up to 35~nm, deposition rates were 6-7 nm/min) and 30 W for the highest ones (up to 100~nm, deposition rates 8-10 nm/min).
In order to clean the 2'' target, 10 minutes of pre-sputtering was performed prior to each deposition. The sample-to-target distance was about 10 cm. Substrates were pre-heated and then kept at a temperature of 600$^\circ$C during deposition. Prior to admittance of Ar gas in the sputtering chamber, up to $p~\sim$ 7$\times$10$^{-3}$ mbar, the pressure rose up to about 3$\times$10$^{-5}$ mbar.
Similar procedures have been followed by other groups for producing single crystalline Ru films\cite{SutterAPL2010,SauerbreyCGD2016,AnemonePhD2017}. A total number of 15 films were grown at 20~W and 35 at 30~W; five of them with thicknesses between 7~nm and 100~nm were analyzed by RBS / channeling. These experiments were performed at the standard chamber of the 5~MV tandem ion accelerator of CMAM using He ions of 1.8 MeV energy~\cite{Redondo2021}. Thickness and composition of the films were determined by comparing experimental RBS spectra in random geometry with simulations performed using the SIMNRA software package~\cite{SIMNRA1999}.
Parameters used in the simulation were: scattering angle: 170.9$^{\circ}$, incidence angle: 0$^{\circ}$ respect surface normal,
Nr. of particles times detector solid angle: 6.0$\times$10$^{11}$ (part)*sr, detector resolution: 20~keV, atomic density of Ru(0001) layers: 1.58$\times$10$^{15}$ (at)cm$^{-2}$.

The surface quality of the Ru films was checked by low-energy electron diffraction (LEED) performed inside an ultra-high vacuum (UHV) chamber, where the samples were cleaned {\em in-situ} after having been transferred at atmospheric pressure from the growth chamber. For this purpose, usual methods for surface preparation inside an UHV chamber (base pressure: 5$\times$10$^{-10}$ mbar) were applied: a cycle of sputtering with 1~keV Ar ions at room temperature (10 min, $p$~=7$\times$10$^{-5}$ mbar) and heating at 700$^\circ$C in oxygen (5 min, $p$~=5$\times$10$^{-7}$ mbar) was enough in order to remove C contamination, followed by a final annealing of 1 min in vacuum up to a temperature of 1200$^\circ$C, whereby the pressure rose up to 5$\times$10$^{-8}$ mbar.

\section{Results and Discussion}

\subsection{Ru(0001) single crystal} 

\begin{figure}
    \centering
    \includegraphics[width=0.50\textwidth]{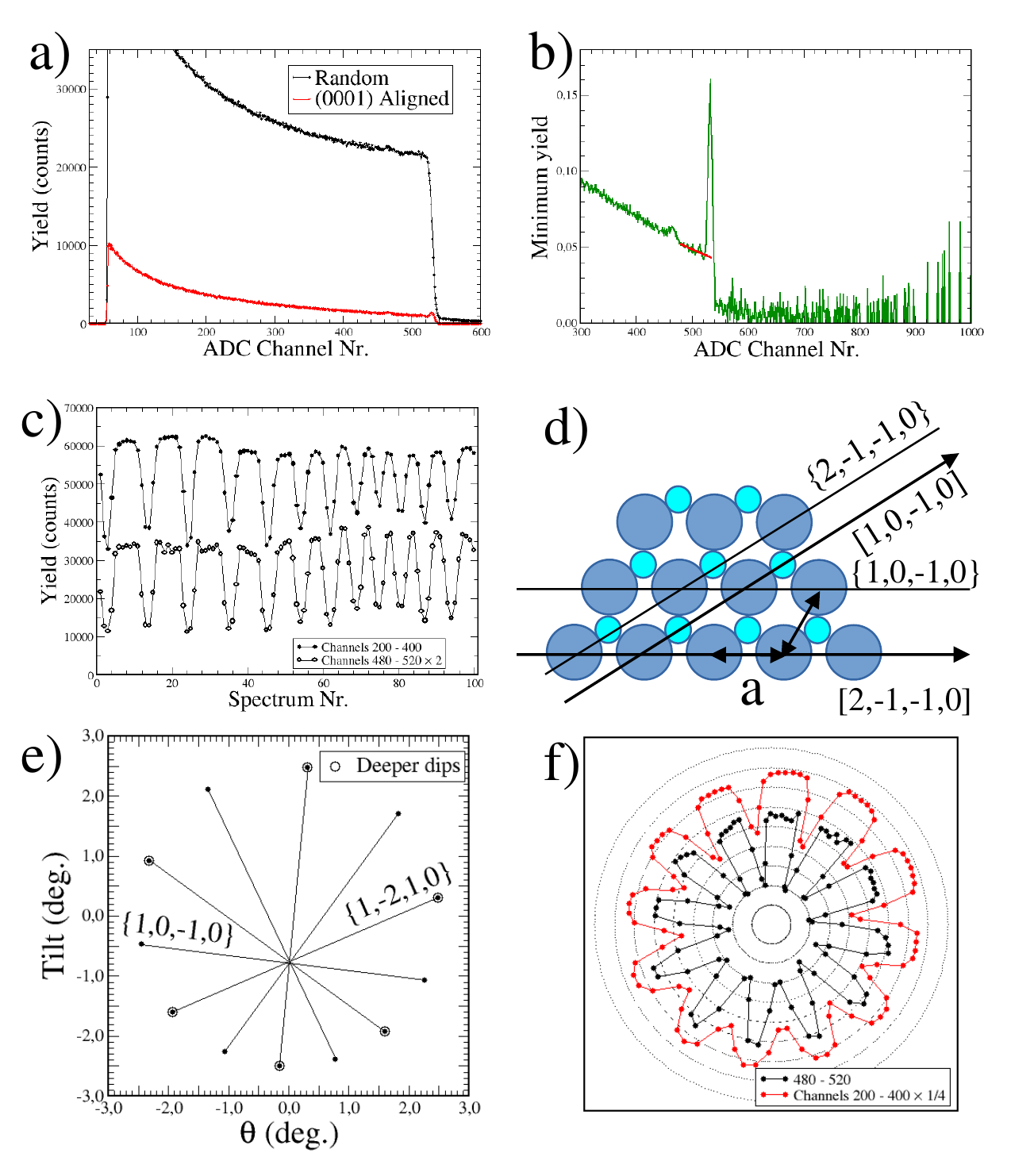}
    \caption{a) RBS spectra of the Ru(0001) single crystal at random and channeling directions of the incoming He beam; b) Minimum yield of the signal calculated as the ratio of the channeling and random signals. The red line is a linear fit in the range of ADC channels (480-520) corresponding to the subsurface region and extrapolated to the surface peak position (channel 533); c) angular spectra in different energy windows; d) schematic {\em hcp} structure. e) $\theta$-Tilt plot of the positions of the minima of the yield; f) Polar plot of the angular spectra.}
    \label{fig:Ru}
\end{figure}

Helium ions backscattering spectra of the Ru(0001) crystal were recorded by varying the sample’s orientation around the vertical axis ($\theta$ angle) and the horizontal transverse axis (Tilt angle) of the goniometer. The corresponding set of directions forms a cone around the sample holder’s normal with a typical opening angle of 2.5 degrees. In this way, if the misalignment of the crystal with respect to the sample holder is small enough, the position of the hexagonal-close-packed ({\it hcp}) [0001] channel will lie inside the cone and the incoming beam will cross the corresponding planar channels around the axial [0001] direction as the angular scan is performed. A reliable spectrum representative for scattering along a “random” direction can be constructed by adding all the spectra measured in the described angular scan. The result for the Ru(0001) crystal is shown in Fig.~\ref{fig:Ru}a). On top of the expected step-like aspect of the Ru signal, only a small hump can be observed around channel numbers close to 465; this is due to Co or Fe atoms in the near-surface region remaining from previous growth experiments of oxides of these transition metals on the Ru crystal and not having been completely removed by high-temperature (~1300~K) flashing. Quantification of the spectrum gives a concentration of 5\% Co/Fe in the topmost 20~nm of the Ru(0001) crystal.

On the “random” spectrum, ranges of analog-to-digital (ADC) channels can be defined, which translate into energy windows, in turn corresponding to ions being scattered by target atoms located at given depths inside the sample. A plot of the backscattering yield integrated inside one such window as a function of the incoming beam direction will show channeling dips as the beam sucessively enters and leaves the directions of planar channeling around the axial {\it hcp} [0001] direction.  An example is shown in Figure~\ref{fig:Ru}c). One can clearly see 12 channeling dips for a complete revolution around the sample holder’s normal direction. Furthermore, the minima alternate according to their depth, which hints to 2 different families of crystallographic planes, the most compact ones perpendicular to {\it hcp} (0001), i.e., the $\{10\overline{1}0\}$ and the $\{1\overline{2}10\}$ planes. The fact that the intensity distribution is not completely symmetrical is due to the small misalignment of the crystallographic [0001] direction with respect to the sample holder’s normal.

The angular positions of the yield minima can be used to locate the exact position of the [0001] axis by plotting them in a ($\theta$-Tilt) graph as shown in Fig.~\ref{fig:Ru}e). Connecting with straight lines the two minima corresponding to the same crystallographic plane, all the lines must cross at one point, within the tolerance given by the precision of the goniometer motion. This gives us the direction [0001] of axial channeling and allows us also to correct for the crystal misalignment with respect to the sample holder and plot the corresponding yield as a function of the azimuthal angle perpendicular to the [0001] direction in a polar plot, as done in Fig.~\ref{fig:Ru}f) for two different energy windows.
Knowledge of the channeling [0001] direction allows also to record a backscattering spectrum with incoming ions aligned along this direction; This is included in Fig.~\ref{fig:Ru}a). A total ion fluence of 20 $\mu$C has been used, equal to the sum of the fluences of all the angular spectra. A reduced yield in comparison with the random direction is clearly visible, as well as the presence of a surface peak.

The {\it minimum yield}, defined as the ratio between the scattering yields for the channeling and the random directions, can be used to assess the quality of the crystal. This ratio is calculated and plotted in Fig.~\ref{fig:Ru}b). Extrapolating linearly to the position of the surface peak gives a ratio of 4.4 \%, a value in the typical range of good-quality single crystals~\cite{SiMBEII2018}.

The planar channeling directions can be identified by comparing their minimum yields.
For the $\{1\overline{2}10\}$ planes, the interplanar distance is $d$ = $a$/2. As shown in Figure~\ref{fig:Ru}d), the width of the region blocked by the crystal atoms is 2$b$, where $b$ is of the order of the mean thermal vibration amplitude, which can be estimated for Ru with Debye’s model to be about 0.1 \AA{} at room temperature. Therefore, from this geometrical estimation, a minimum yield for planar channeling can be calculated by $\chi_{min} = 2b / (0.5 a)$. Analogously, for the $\{10\overline{1}0\}$ planes, which contain the $\langle1\overline{2}10\rangle$ directions in which nearest-neighbor Ru atoms are aligned, the corresponding interplanar distance is $d = (\sqrt{3}/2) a$. But inside this region, a second plane of Ru atoms is located, so that the width of the blocked region amounts to 4$b$. So we have $\chi_{min} = 4b / (0.87 a)$, so that the ratio of minimum yields amounts to 0.87 in our case. We therefore conclude that the planes giving the lower minimum yield are the $\{1\overline{2}10\}$ ones. Furthermore, in good agreement with this value, we obtain experimentally an average ratio of 0.88 between the two non-equivalent families of lattice planes, for the energy window corresponding to the near-surface region (ADC channels 480-520).

\subsection{Ru/Al$_2$O$_3$(0001) films}

\begin{figure}
    \centering
    \includegraphics[width=0.5\textwidth]{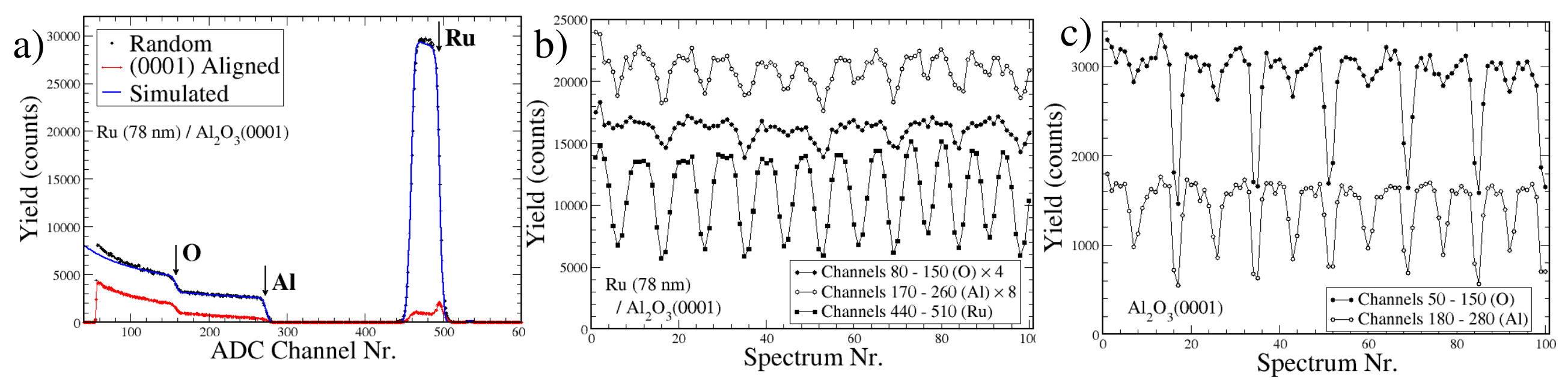}
    \caption{a) RBS spectra of a a 78~nm Ru/Al$_2$O$_3$(0001) film at random and channeling directions of the incoming He beam; 
    b) angular spectra of the film in different ADC windows; c) angular spectra of the film's Al$_2$O$_3$(0001) substrate protected by a mask from Ru deposition.}
    \label{fig:Ru78}
\end{figure}

Several Ru films were grown with thicknesses ranging from 7 nm to 100 nm on Al$_2$O$_3$(0001) single crystals. They were analyzed by means of RBS/Channeling. The same procedures as described above for the case of the Ru(0001) crystal were applied for the recording of the angular spectra, for the construction of the random spectrum and for the determination of the direction of axial channeling of the Ru films. As a typical example, we consider first a film of 78 nm thickness. Figure~\ref{fig:Ru78}a) shows RBS spectra in random as well as in channeling configurations. It includes the result of a SIMNRA simulation, which reproduces well the random spectrum and allows the determination of the film thickness.

In Fig.~\ref{fig:Ru78}a), the signals corresponding to ions scattering at the Ru atoms in the film and to Al and O atoms in the substrate can be clearly observed at decreasing energies. In addition, there is a small peak visible above the Ru signal, at around channel number 535. This is due to a small W contamination coming from the filament, as shown by the quantification of the spectrum giving a W concentration of 0.9\% in the film. The peak also appears in the spectra of the rest of the films analyzed; the W concentrations determined were always below this value.

An important reduction of the yield is observed in axial channeling for all elements, in particular for ions scattering at the Ru film, which hints at a good crystalline quality of the film. A clear surface peak is observed, corresponding to ions scattering at the surface layers, which are not efficiently shadowed by crystal atoms above and therefore experience a uniform flux of incoming particles. Remarkably, there is no interface peak at the low-energy end of the Ru signal, which points to a well-ordered interface between film and substrate.

\begin{figure}
    \centering
    \includegraphics[width=0.5\textwidth]{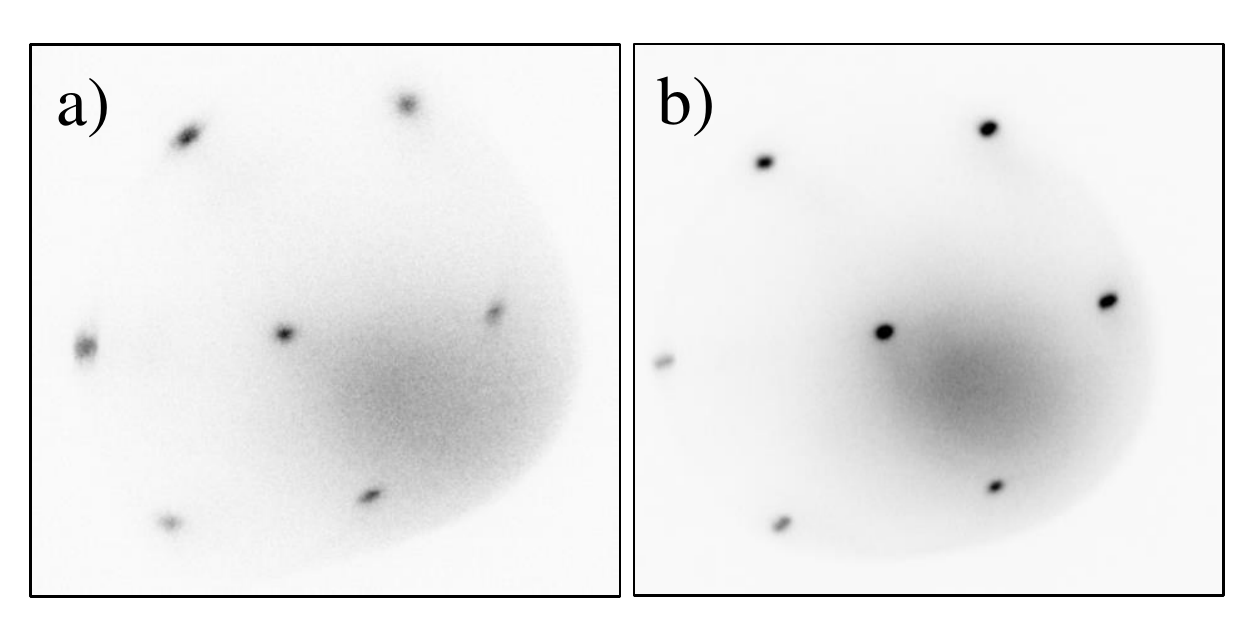}
   \caption{a) LEED patterns an electron energy of 45~eV of (a) an 80~nm Ru/Al$_2$O$_3$(0001) film (appears slightly distorted due to a small misalignment in the electron optics) and (b) a 15~nm Ru/Al$_2$O$_3$(0001) film. Both patterns have been obtained after preparing the respective surfaces according to the procedure described in the experimental section.}
    \label{fig:LEED}
\end{figure}

The reduction of yield can be quantified here also by the minimum yield, i.e., the ratio of the channeling to random yields. This value, calculated for the ADC channels 470-485 corresponding to the film and extrapolated to the position of the surface peak, amounts to 2.7 \%, which indicates an excellent crystalline quality of the film and a good epitaxial orientation of the [0001] directions of the {\em hcp} Ru lattice and the {\em corundum} Al$_2$O$_3$ structure. This is further confirmed by calculating the minimum yield of the signal corresponding to the substrate, i.e. of the ions scattering at Al atoms, as the ratio of the yields at the channeling and random geometries, extrapolated to the onset of the substrate Al signal. This amounts to 15 \% for the 78 nm film, which shows a small degree of dechanneling experienced by the ions in their passage through the film and confirms its high epitaxial quality. In addition, the surface of the films can be prepared in UHV following the procedures described in the experimental section, in order to remove surface contaminants (adsorbed gases, surface oxide layers, carbon segregated from the bulk, etc.) and improve surface ordering. 
In particular, surface C produces very characteristic LEED patterns (moiré superstructures) on Ru(0001) due to the formation of graphene overlayers~\cite{MoritzPRL2010}. Our films gave (2$\times$2) LEED patterns after oxygen exposure, due to O adsorption on Ru(0001) and (1$\times$1) patterns after flashing to high temperatures in UHV, as shown in 
Fig.~\ref{fig:LEED}a for an 80~nm Ru film, implying that it 
exposes a clean and well ordered Ru(0001) surface.

Angular scans of the yield integrated in different energy windows are shown in Fig.~\ref{fig:Ru78}b. Twelve dips are clearly observed for the Ru window, of alternating depths, as in the case of the Ru(0001) single crystal. In addition, also 12 alternating dips are present at the Al signal, while for the O signal 6 dips are clearly visible and further 6 ones are less pronounced. Based on the detailed work of Wu et al.\cite{WuNIMB} for AlGaN on Al$_2$O$_3$(0001), we can assign the deeper minima to the $\{10\overline{1}0\}$ planes and the shallower ones to the $\{1\overline{2}10\}$ planes of Al$_2$O$_3$(0001). In Fig.~\ref{fig:Ru78}c, the result for the clean substrate of this particular, 78 nm thick film is included, which shows a similar relation between the depths of the channeling dips. This was recorded in a region protected by a mask from Ru deposition, so that the epitaxial relation between film and substrate can be unambiguously determined. 

It can be clearly appreciated in Fig.~\ref{fig:Ru78}c) that the angular positions of the deeper planar channels of the Ru lattice coincide with the deeper ones of the substrate. This implies that the $\{1\overline{2}10\}$ planes of Ru(0001) are aligned parallel to the $\{10\overline{1}0\}$ planes of Al$_2$O$_3$(0001). This was already shown in Refs.~\cite{YamadaJJAP2002,MilosevicJAP2018} and is in fact to be expected, since between the $<1\overline{2}10>$ directions of the Ru hcp lattice (along which nearest-neighbors, close-packed Ru atoms are aligned, at a nearest-neighbor distance of $a_{Ru}$ = 2.706~\AA) and the $<10\overline{1}0>$ directions of the sapphire lattice (along which nearly close packed O atoms are aligned, at a nearest-neighbor distance of $a_{sapp}/\sqrt{3}$ = 2.747~\AA), the mismatch amounts only to -1.5~\%. An analogous epitaxial relation has been determined for AlGaN(0001) films (wurtzite structure) on Al$_2$O$_3$(0001)\cite{WuNIMB}.
Comparing with related epitaxial metallic systems grown by sputtering, Pd(111)/Al$_2$O$_3$(0001) films show two different in-plane orientations~\cite{Aleman2018}, while Zr(0001)/Al$_2$O$_3$(0001) only one~\cite{Fankhauser2016}, similar to Pt(111)/Al$_2$O$_3$(0001) when a mixture of O$_2$ and Ar gases is employed~\cite{Tanaka2017}.

\begin{figure}
    \centering
    \includegraphics[width=0.5\textwidth]{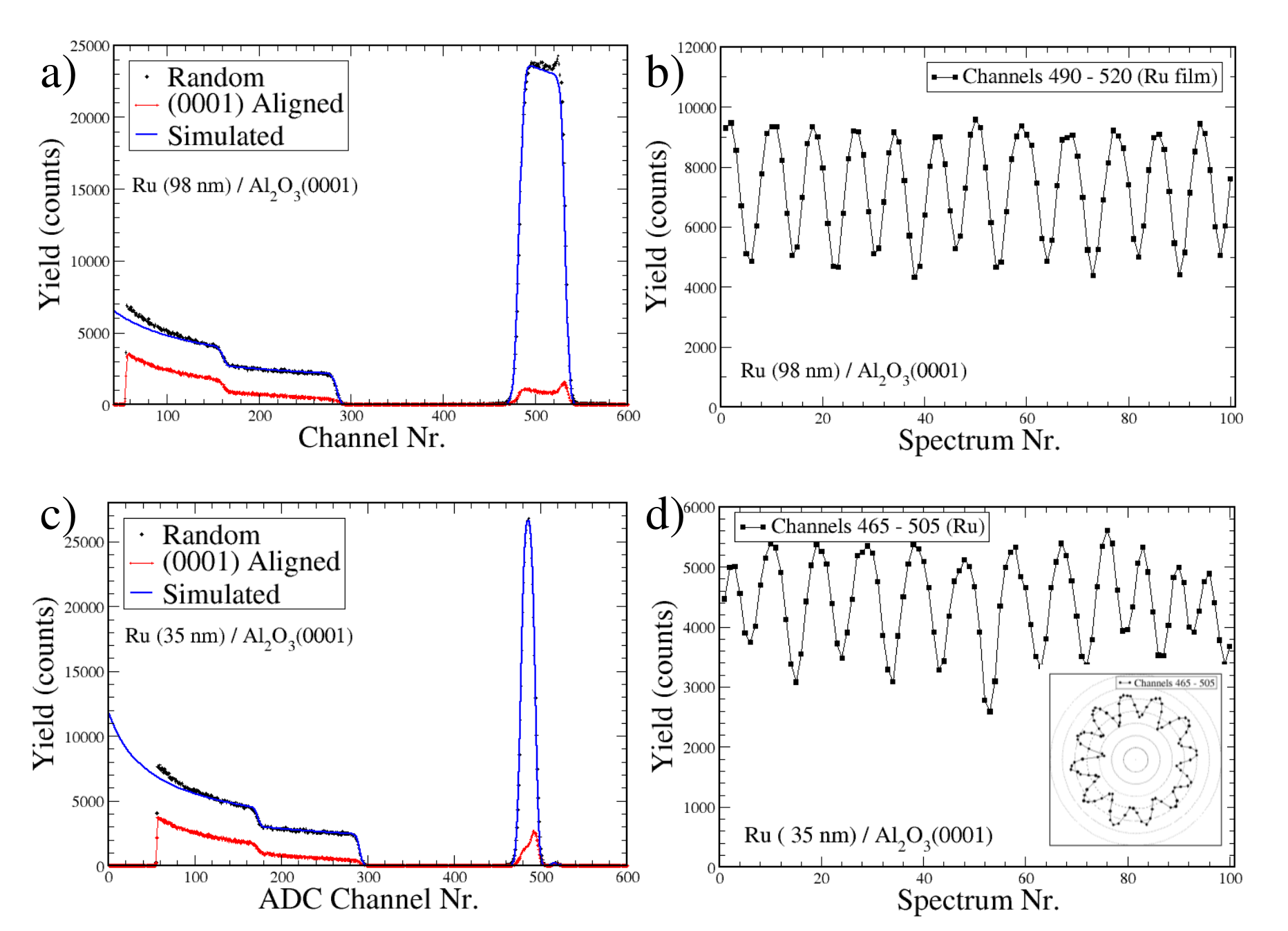}
    \caption{a) RBS spectra at random and channeling incidence directions of the 98~nm Ru film; b) corresponding angular spectra integrated in the given ADC channels window, corresponding to ions backscattered at film's atoms; c) RBS spectra at random and channeling incidence directions of the 35~nm Ru film (channel numbers cannot be directly compared with those in a) because they were measured on different days with different detector settings) ; d) corresponding angular spectra. Insert: polar plot of the backscattered signal of the 35~nm film.}
    \label{fig:Ru98_35}
\end{figure}

Similar results were obtained for thicker (98 nm) as well as for thinner films (35 nm). Figures~\ref{fig:Ru98_35}a) and~c) show random and channeling backscattering spectra for a 98 nm and a 35 nm Ru / Al$_2$O$_3$(0001) film, respectively, together with the corresponding simulations. The minimum yields extrapolated to the surface of the films amount to 2.3 \% for the 100 nm film and to 3.1 \% for the 35 nm film, respectively, which shows the good epitaxial character of both films. In addition, the minimum yield of the signals corresponding to the substrate (ions scattering at Al atoms), extrapolated to the interface, amounts to 17 \% for both the 98 nm and 35 nm films, respectively, showing also here a limited dechanneling of the ions passing through the films and confirming their high crystallinity. Figures~\ref{fig:Ru98_35}b) and~d) show the respective angular scans of the backscattered intensity, integrated in the windows corresponding to the signal of the films, showing the expected 12 dips in both cases. The ratios of minimum yields between the ones corresponding to the $<1\overline{2}10>$ and $<10\overline{1}0>$ families of lattice planes amount to 0.90 and 0.86, for the 98 nm and 35 nm films, respectively, as compared to the geometrically expected 0.87. We therefore conclude that films of thicknesses in the range 35 – 100 nm grow epitaxially on Al$_2$O$_3$(0001) with good quality.
The insert in Fig.~\ref{fig:Ru98_35}d) shows a polar plot of the 35 nm - Ru film signal.

\begin{figure}
    \centering
    \includegraphics[width=0.5\textwidth]{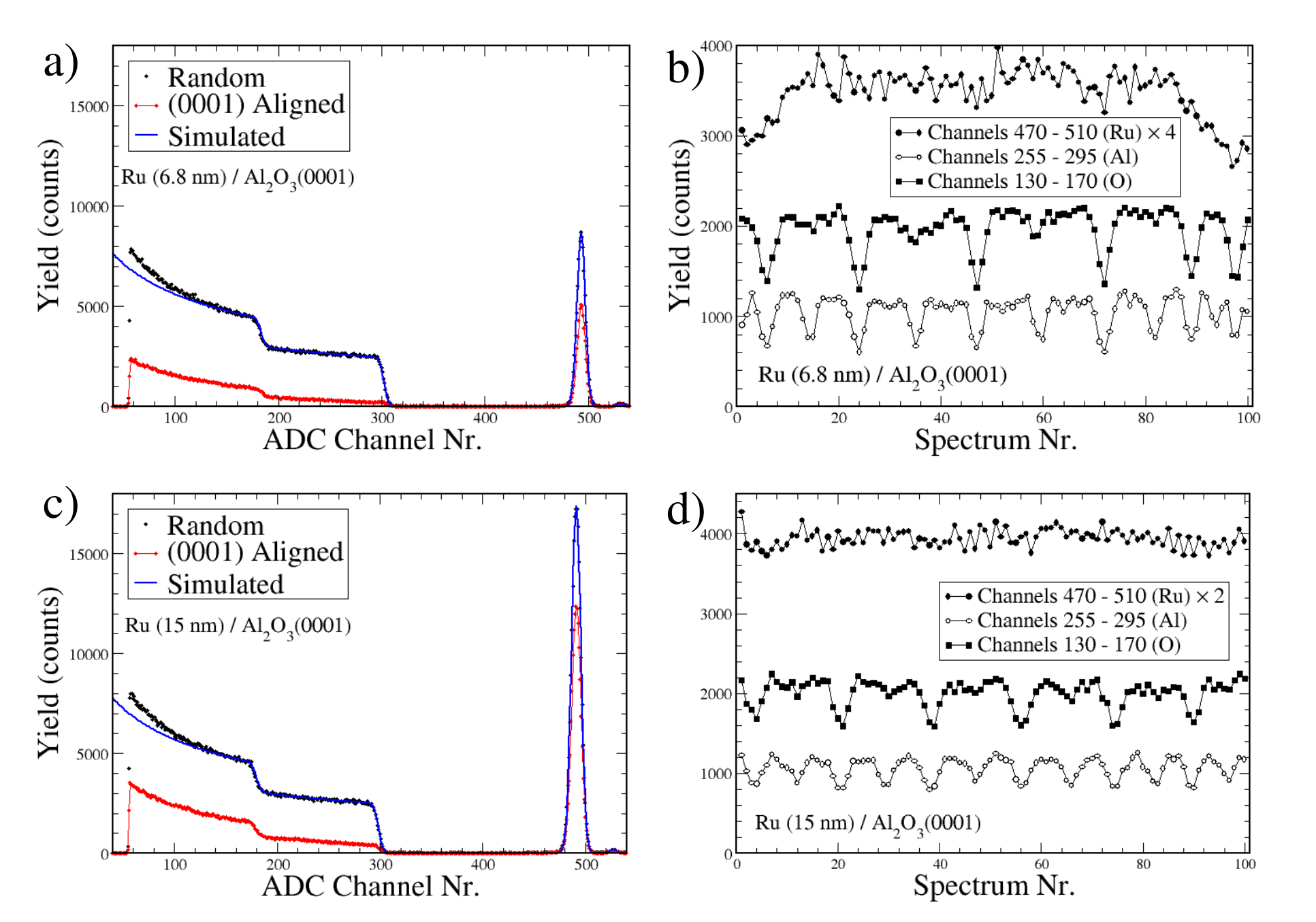}
    \caption{a) RBS spectra at random and channeling incoming directions of the 6.8~nm film; b) corresponding angular spectra; c) RBS spectra at random and channeling incoming directions of the 15~nm film; d) corresponding angular spectra.}
    \label{fig:Ru15_6.8}
\end{figure}

We consider now thinner films. Figures~\ref{fig:Ru15_6.8}a)
and c) show backscattering spectra for a 6.8 nm and a 15 nm and film, respectively, both in channeling and in random geometries, as well as the simulated spectra for the random cases. The first observation is that both films are so thin that that the extended shape of the Ru signal characteristic of a film is not yet developed and there is only a single, asymmetric peak visible.

On these random spectra, energy windows corresponding to ions scattered at Ru atoms of the films and at Al and O atoms of the substrates have been defined. The corresponding backscattering yields integrated in these windows have been plotted in Figures~\ref{fig:Ru15_6.8}b) and d) as a function of the incoming beam direction. The result is that no appreciable planar channeling is observed for the Ru films, presumably because they are so thin that practically all the backscattered signal is coming from the surface region where atoms have not enough neighbors above to protect them from the incoming ion flux. However, the channeling dips can be clearly observed and identified for the signals corresponding to the sapphire substrates, as in the cases of the thicker films. This fact allows us, following the same procedure applied for the thicker films, to determine the orientation of the axial [0001] channel and record the corresponding channeling spectra for the films. These are included in Figures~\ref{fig:Ru15_6.8}a) and c).

We can appreciate that there is a significant reduction of the yield in the axial channeling configuration. In order to quantify this reduction, we have to note that the procedure of extrapolating the minimum yield of the film to the surface, as done for the thicker films and the Ru crystal, does not work here because there is practically no signal from the film apart from that of the surface itself. Therefore, the minimum yield values obtained by building the ratio of the signals at channeling and random configurations are not directly comparable with those of the thicker films. The results for the Ru signal of the films are 72 \% and 58 \% for the 15 nm and the 6.8 nm films, respectively. On the other hand, since the channels can be well observed for the substrate signals, the dechanneling effect produced by the film on the incoming ion flux can be estimated and compared with the results for the thicker films. 

\begin{table}
\centering
\begin{tabular}{c|c|c}
Thickness & $\chi_{min}^{surf}$ & $\chi_{min}^{interf}$ \\\hline
Ru(0001) & 4.4 \% & - \\
98 nm & 2.3 \% & 17 \% \\
78 nm & 2.7 \% & 15 \% \\
35 nm & 3.1 \% & 17 \% \\
15 nm & 72 \% (*) & 15 \% \\
6.8 nm & 58 \% (*) & 7.2 \% \\
Al$_2$O$_3$(0001) & - & 0.76 \% \\
\end{tabular}
\caption{\label{tab:yields} Compilation of the minimum yields of the samples analyzed in this work: Ru(0001) single crystal,  Ru/Al$_2$O$_3$(0001) films of different thicknesses and Al$_2$O$_3$(0001) substrate.
$\chi_{min}^{surf}$ is the minimum yield of the Ru signal extrapolated to the position of the surface peak (when possible, exceptions are the two thinnest films, marked with~*) and $\chi_{min}^{interf}$ is the minimum yield of the substrate, extrapolated to the onset of the Al signal at the interface with the film or to the surface of the substrate (see text).}

\end{table}

For the Al signal of the substrate, the minimum yield extrapolated to the interface amounts to 15~\% for the 15~nm film and to 7.2~\% for the 6.8~nm film.
While the value for the latter is by a factor of 2 smaller than those obtained for thicker films, a fact attributable to its smaller thickness, the 15 \% obtained for the 15 nm film is quite similar to the values obtained for the films of thickneses of 35 nm and higher. For the clean Al$_2$O$_3$(0001) substrate, the minimum yield is a very small 0.76 \%, extrapolated to the surface.
The results for the minimum yields of the films analyzed in this work are shown in Table~\ref{tab:yields}
together with those for the Ru(0001) and Al$_2$O$_3$(0001) substrates.
The fact that for the thinnest film analyzed (6.8~nm) the minimum yield is half that of the next, twice as large thicker film (15~nm), points to a similar defect concentration in both films. Furthermore, for thicknesses higher than 15~nm, the minimum yield does not increase anymore, but stays constant at 15 \% - 17 \%, which shows that the amount of defects saturates.  
This can be rationalized in the following way: in a low mismatch system like Ru/Al$_2$O$_3$(0001), very thin films grow pseudomorphically with the substrate; at a given thickness, the introduction of misfit dislocations (MD's) starts to relax the strain until bulk lattice parameters are achieved. In our system, at 6.8~nm thickness the introduction of MD's has already begun, while at 15~nm, the transition is practically complete; from this thickness on, films grow with lattice parameters relaxed to the bulk values (XRD shows that 78~nm films have a bulk lattice constant~\cite{MilosevicJAP2018}).
The surface of the thinnest films can be prepared in a similarly clean and ordered state as for the thicker films, as shown by the LEED pattern of a 15~nm Ru film (Fig.~\ref{fig:LEED}b).

\section{Summary}

In conclusion, we have grown Ru films on Al$_2$O$_3$(0001) substrates by magnetron sputtering, with thicknesses ranging between 7 nm and 100 nm. A careful characterization has been performed by RBS / channeling measurements. We have shown that films in the range 35 nm – 100 nm grow epitaxially with relations Ru(0001) $\|$ Al$_2$O$_3$(0001) and
Ru$\langle1\overline{2}10\rangle$
$\|$ Al$_2$O$_3\langle10\overline{1}0\rangle$, with a very good crystalline quality, as shown by the observation of axial and planar channeling and by the values obtained for the minimum yield of the order of 2 \% - 3 \%  for axial channeling vs. random incidence, quite comparable to those obtained for a reference Ru(0001) single crystal. 
No evidence of interfacial disorder is found.
Thinner films in the range 7 nm – 15 nm do not develop the characteristics of the bulk material as to be fully quantified by RBS channeling; for those, relaxation of epitaxial strain is not complete, 
although they produce a similar degree of dechanneling as the thicker films. 
The surfaces of Ru(0001) / Al$_2$O$_3$(0001) thin films can be prepared in a clean and ordered state, as checked by surface-sensitive techniques like LEED, to allow further epitaxial growth of films or nanostructures on top.

\section{Aknowledgments}

This research has been supported by 
grant RTI2018-095303-B-C51 funded by MCIN/AEI/10.13039/501100011033 and by “ERDF A way of making Europe”, and by grant  S2018-NMT-4321 funded by the Comunidad de Madrid and by “ERDF A way of making Europe”.
We acknowledge the support from CMAM for the beamtime proposals with codes STD009/20, STD021/20 and STD027/20.

\bibliographystyle{elsarticle-num}
\bibliography{RBS_Ru}

\end{document}